\documentclass[conference]{IEEEtran}
\IEEEoverridecommandlockouts
\usepackage{cite}
\usepackage{amsmath,amssymb,amsfonts}
\usepackage{algorithmic}
\usepackage{graphicx}
\usepackage{textcomp}
\usepackage{xcolor}
\usepackage{subfigure}
\usepackage{booktabs}
\usepackage{multirow}
\usepackage{array}

\def\BibTeX{{\rm B\kern-.05em{\sc i\kern-.025em b}\kern-.08em
    T\kern-.1667em\lower.7ex\hbox{E}\kern-.125emX}}
\newcommand\blfootnote[1]{%
  \begingroup
  \renewcommand\thefootnote{}\footnote{#1}%
  \addtocounter{footnote}{-1}%
  \endgroup
}

\DeclareRobustCommand*{\IEEEauthorrefmark}[1]{%
\raisebox{0pt}[0pt][0pt]{\textsuperscript{\footnotesize\ensuremath{#1}}}}

\begin{document}

\title{On the Rate-Distortion-Complexity Trade-offs \\ of Neural Video Coding}

\author{%
\IEEEauthorblockN{
Yi-Hsin Chen\IEEEauthorrefmark{1} 
\quad Kuan-Wei Ho\IEEEauthorrefmark{1} 
\quad Martin Benjak\IEEEauthorrefmark{2}
\quad Jörn Ostermann\IEEEauthorrefmark{2}
\quad Wen-Hsiao Peng\IEEEauthorrefmark{1}
}
\IEEEauthorblockA{%
\IEEEauthorrefmark{1}National Yang Ming Chiao Tung University, Taiwan \quad 
\IEEEauthorrefmark{2}Leibniz Universität Hannover, Germany}
}


\maketitle

\begin{abstract}
This paper aims to delve into the rate-distortion-complexity trade-offs of modern neural video coding. Recent years have witnessed much research effort being focused on exploring the full potential of neural video coding. Conditional autoencoders have emerged as the mainstream approach to efficient neural video coding. The central theme of conditional autoencoders is to leverage both spatial and temporal information for better conditional coding. However, a recent study indicates that conditional coding may suffer from information bottlenecks, potentially performing worse than traditional residual coding. To address this issue, recent conditional coding methods incorporate a large number of high-resolution features as the condition signal, leading to a considerable increase in the number of multiply–accumulate operations, memory footprint, and model size. Taking DCVC as the common code base, we investigate how the newly proposed conditional residual coding, an emerging new school of thought, and its variants may strike a better balance among rate, distortion, and complexity.


\end{abstract}

\begin{IEEEkeywords}
Learned video compression, conditional coding, and conditional residual coding.
\end{IEEEkeywords}  

\vspace{-0.3cm}
\blfootnote{This work is supported by National Science and Technology Council, Taiwan (112-2634-F-A49-007-, 110-2221-E-A49-065-MY3, 111-2923-E-A49-007-MY3), and National Center for High-performance Computing, Taiwan.}

\section{Introduction}
\label{sec:intro}
Learned video compression~\cite{dcvc,canf,tcm,hem,maskCRT,dcvc_dc,dcvc_fm} holds the promise of revolutionizing the way high-efficiency video compression systems are developed. The state-of-the-art methods~\cite{hem, maskCRT} have shown comparable coding performance to VVC~\cite{vvc} under the low-delay configuration. Some~\cite{dcvc_dc, dcvc_fm} even outpace ECM~\cite{ecm}. Different from traditional codecs~\cite{hevc,vtm,ecm}, which mostly adopt residual coding to encode the prediction residue $x_t-x_c$ between a target frame $x_t$ and its temporal predictor $x_c$, modern neural video codecs~\cite{dcvc,tcm,hem,dcvc_dc,dcvc_fm} attribute part of their success to the non-linear exploitation of the temporal predictor $x_c$. Specifically, $x_c$ is no longer used in the pixel domain to compute the prediction residue, but instead, it is utilized to condition the learned inter-frame codec in coding the target frame $x_t$. This technique is widely known as conditional coding. Theoretically, the conditional entropy $H(x_t|x_c)$ is less than or equal to the residual entropy $H(x_t-x_c)$~\cite{mmsp, pcs22}. 

Although showing promising coding performance, \cite{pcs22,cond_res_coding} indicate that conditional coding may suffer from the information bottleneck issue. That is, the information from the temporal predictor $x_c$ may be lost during the feature extraction process in formulating a condition signal. As a result, in some cases, conditional coding may perform worse than residual coding. To mitigate the effect of this information bottleneck, a general trend in the state-of-the-art conditional coding approaches~\cite{dcvc,tcm,hem,dcvc_dc,dcvc_fm} is to extract a large number of high-resolution features from $x_c$. However, this results in a considerable increase in the number of multiply–accumulate operations, memory footprint, and model size.

\begin{figure}[t]
    \centering
    \includegraphics[width=1\linewidth]{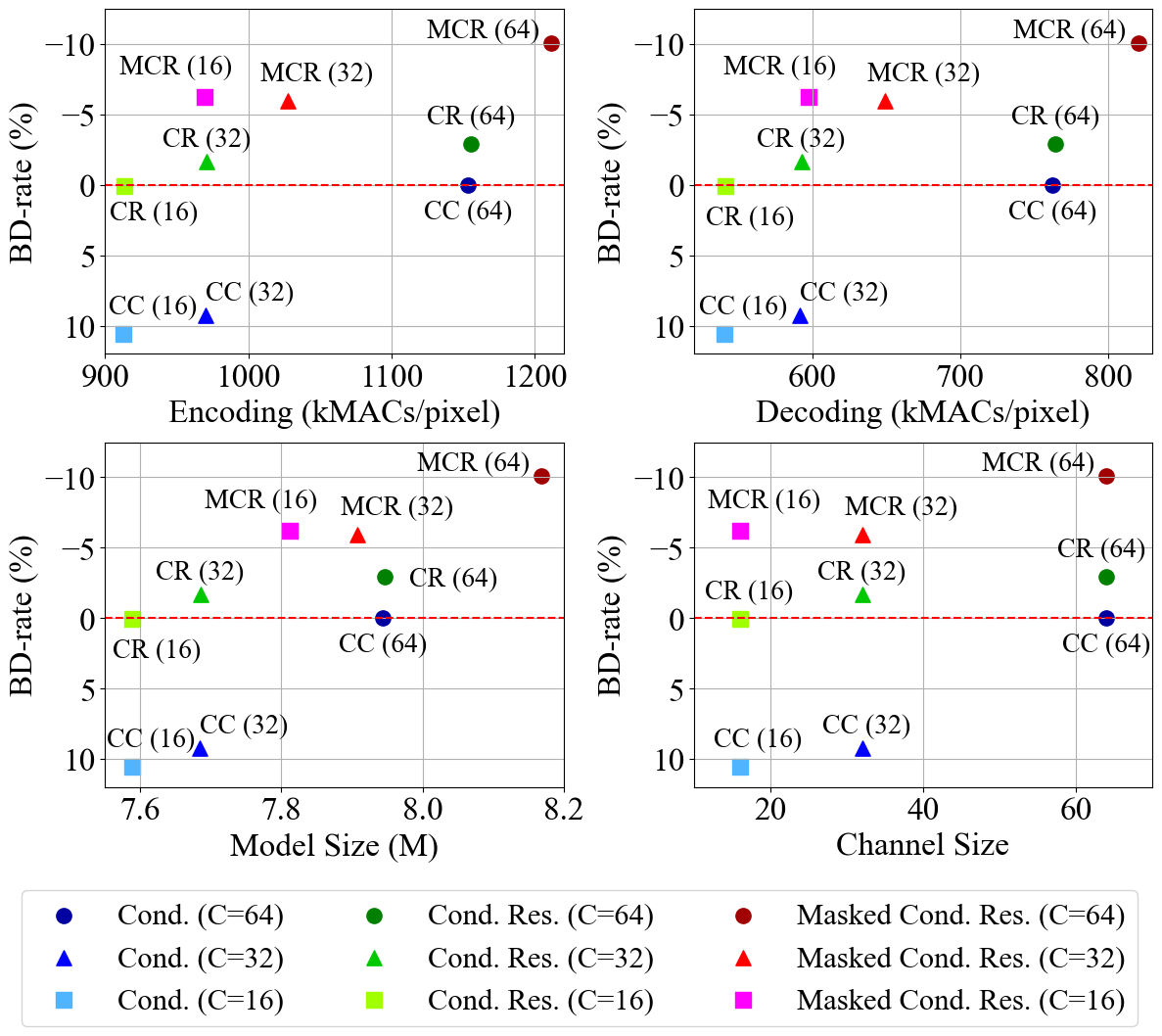}
    \vspace{-0.6cm}
    \caption{BD-rate versus complexity (measured in the encoding kMACs/pixel, decoding kMACs/pixel, model size, and channel size of the full-resolution condition signal for inter-frame coding). CC, CR, MCR in the figure represent conditional coding, conditional residual coding, and masked conditional residual coding, respectively. The vertical axis represents BD-rate in terms of PSNR-RGB. The anchor is conditional coding with 64 full-resolution feature maps as the condition signal. Positive and negative BD-rate numbers indicate rate inflation and reduction, respectively. See Section~\ref{sec:experiment} for more results and evaluation setup.
    }
    \label{fig:teaser}
    \vspace{-0.4cm}
\end{figure}
\begin{figure*}[t]
    \centering
    \includegraphics[width=1\linewidth]{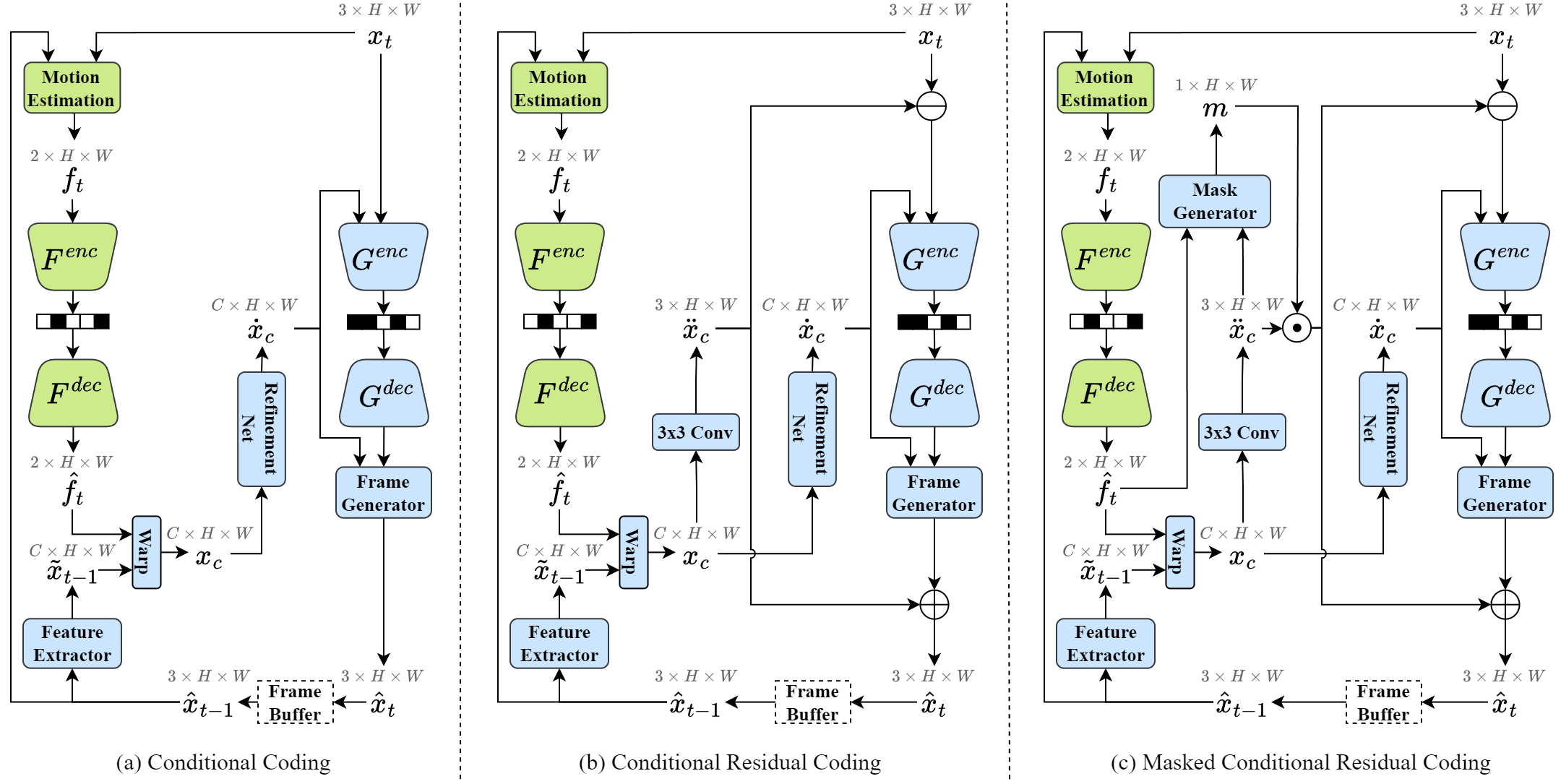}
    \vspace{-0.5cm}
    \caption{The system overview of (a) DCVC~\cite{dcvc} and its variants with (b) conditional residual coding and (c) masked conditional residual coding. In these three variants, we follow DCVC~\cite{dcvc} to incorporate a temporal prior into the entropy model of the inter-frame codec. For brevity, it is not depicted in the figure.}
    \vspace{-0.3cm}
    \label{fig:schemes}
\end{figure*}

To remedy the information bottleneck, Brand~\emph{et al.}~\cite{cond_res_coding} recently introduced conditional residual coding, which encodes the prediction residue $x_t-x_c$ with a conditional inter-frame codec. Brand~\emph{et al.}~\cite{cond_res_coding} show that in theory, the coding efficiency of conditional residual coding is at least as good as that of conditional coding and is less susceptible to the information bottleneck. Ideally, conditional residual coding is able to achieve higher coding performance with lower complexity. However, the theory was validated under a simplified setup with a simple learned video codec. Moreover, the results from \cite{cond_res_coding} are based on the assumption that the temporal predictor has good quality and the entropy of the residue $x_t-x_c$ is less than the entropy of the target frame, i.e., $H(x_t-x_c) \leq H(x_t)$.

Considering that the assumption $H(x_t-x_c)\leq H(x_t)$ of conditional residual coding may be violated in regions with dis-occlusion or unreliable motion estimates, Chen~\emph{et al.}~\cite{maskCRT} propose a masked conditional residual coding scheme, which improves on conditional residual coding by introducing a pixel-wise soft mask to switch between conditional coding and conditional residual coding. While this approach shows promising results, the trade-offs between coding performance and complexity among conditional coding, conditional residual coding, and masked conditional residual coding are not fully discussed in~\cite{maskCRT}. Moreover, the experiments in~\cite{maskCRT} are conducted using a Transformer-based video codec, which differs from the more commonly used CNN-based codecs.

In this work, we use DCVC~\cite{dcvc} -- a typical CNN-based conditional video codec -- as the common code base to explore how the recently proposed conditional residual coding and its variants could potentially strike a better balance among rate, distortion, and complexity. As shown in Fig.~\ref{fig:teaser}, both conditional residual coding and masked conditional residual coding outperform conditional coding in terms of coding performance while maintaining lower complexity. The  main contributions of this work include (1) extending DCVC from conditional coding to conditional residual coding and masked conditional residual coding, and (2) conducting extensive experiments to study the rate-distortion-complexity trade-offs of these coding schemes under a nearly identical framework.

\begin{figure}[t]
    \centering
    \includegraphics[width=\linewidth]{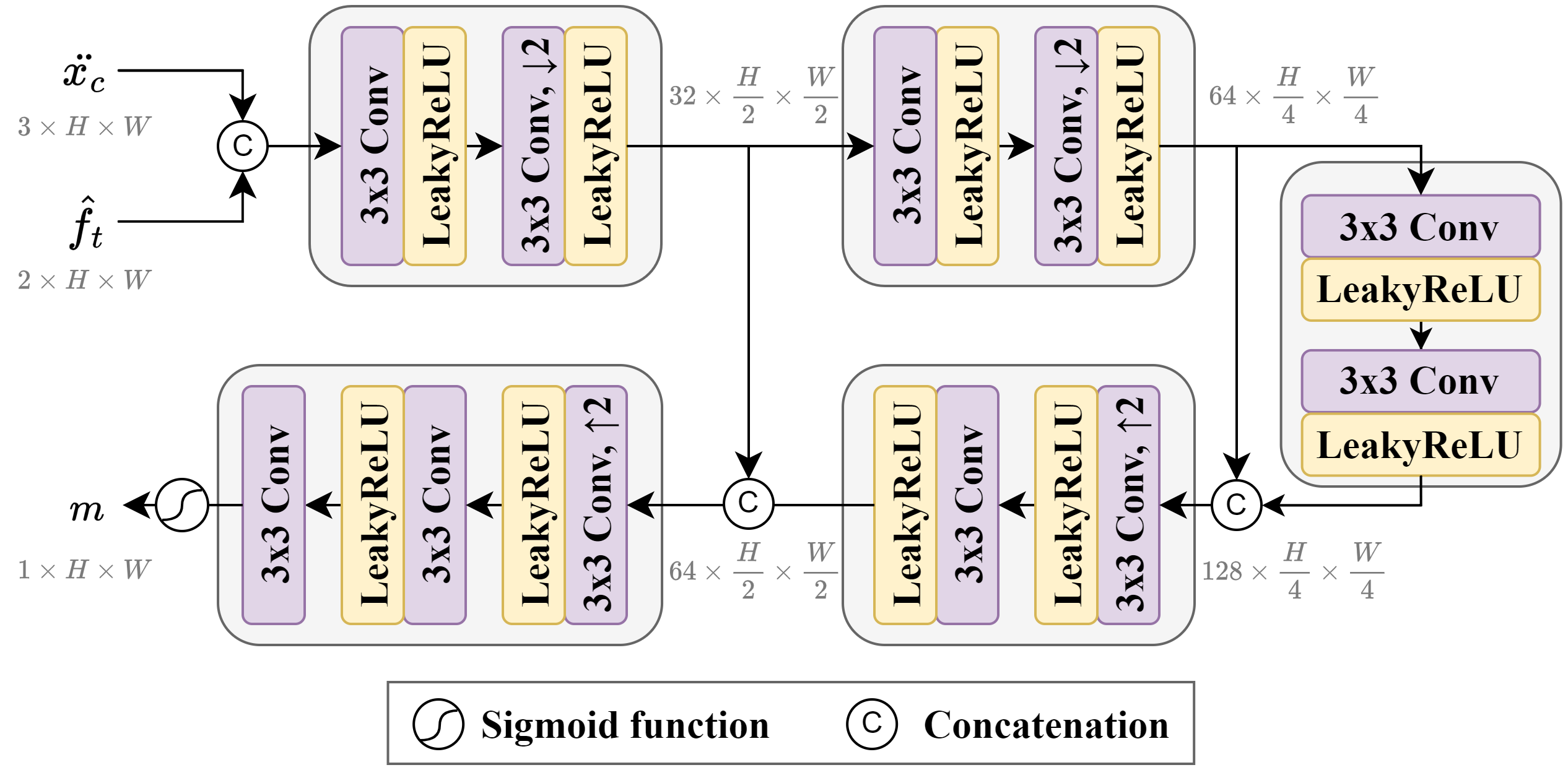}
    \vspace{-0.5cm}
    \caption{The network architecture of the mask generator.}
    \vspace{-0.3cm}
    \label{fig:mask_gen}
\end{figure}
\section{Proposed Method}
\label{sec:method}

\begin{table*}[t]

\centering
\caption{Training procedure. MENet, FE, 3x3 Conv represent the motion estimation network, feature extractor, and 3x3 Conv (only for conditional residual coding and masked conditional residual coding) in Fig.~\ref{fig:schemes}, respectively. The inter-frame codec in this table includes $\{G^{enc}, G^{dec}\}$, the refinement net, and the frame generator in Fig.~\ref{fig:schemes}. EPA stands for the error propagation aware training in \cite{EPA}. $R^{motion}_t$ and $R_t$ denote the estimated motion bit rate and total bit rate, respectively. $D$ measures the mean-squared error between the two signals. $\lambda$ is a hyper parameter.}
\setlength\tabcolsep{2.8pt}
\label{table:training}
\begin{tabular}{cccccccccc}
\hline
\multirow{2}{*}{Phase}               & \multirow{2}{*}{Step} & \multicolumn{5}{c}{Training Modules}                                       & \multirow{2}{*}{Training Objective}               & \multirow{2}{*}{\begin{tabular}[c]{@{}c@{}}\# of \\ Frames\end{tabular}} & \multirow{2}{*}{Epoch} \\ \cline{3-7}
                                     &                       & Inter-frame Codec & MENet & Motion Codec & FE \& 3x3 Conv & Mask Generator &                                                   &                                                                          &                        \\ \hline
Motion Compensation                  & 1                     &                   &       &              & v              &                & $R^{motion}_t + \lambda \times D(x_t, \ddot{x}_c)$ & 3                                                                        & 2                      \\ \hline
\multirow{2}{*}{Inter-frame Coding}  & 2                     & v                 &       &              & v              & v              & $R_t + \lambda \times D(x_t, \hat{x}_t)$          & 3                                                                        & 4                      \\ \cline{2-10} 
                                     & 3                     & v                 &       &              & v              & v              & $R_t + \lambda \times D(x_t, \hat{x}_t)$          & 5                                                                        & 4                      \\ \hline
\multirow{2}{*}{Finetuning}          & 4                     & v                 &       & v            & v              & v              & $R_t + \lambda \times D(x_t, \hat{x}_t)$          & 3                                                                        & 3                      \\ \cline{2-10} 
                                     & 5                     & v                 &       & v            & v              & v              & $R_t + \lambda \times D(x_t, \hat{x}_t)$          & 5                                                                        & 3                      \\ \hline
\multirow{2}{*}{Finetuning with EPA} & 6                     & v                 &       & v            & v              & v              & $R_t + \lambda \times D(x_t, \hat{x}_t)$          & 5                                                                        & 2                      \\ \cline{2-10} 
                                     & 7                     & v                 & v     & v            & v              & v              & $R_t + \lambda \times D(x_t, \hat{x}_t)$          & 5                                                                        & 2                      \\ \hline
\end{tabular}
\end{table*}

\subsection{System Overview}
Fig.~\ref{fig:schemes} illustrates the coding frameworks for DCVC~\cite{dcvc} and its variants that implement conditional residual coding and masked conditional residual coding, respectively. From the high-level perspective, DCVC~\cite{dcvc} includes a motion-coding module (in green color) and a conditional inter-frame coding module (in blue color). To encode a video frame $x_t \in \mathbb{R}^{3 \times H \times W}$ of width $W$ and height $H$, an optical flow map $f_t \in \mathbb{R}^{2 \times H \times W}$ characterizing the motion between $x_t$ and its reference frame $\hat{x}_{t-1} \in \mathbb{R}^{3 \times H \times W}$ is estimated by a motion estimation network and compressed by the motion codec $\{F^{enc}, F^{dec}\}$. The decoded optical flow map $\hat{f}_t$ is then utilized to warp the features $\tilde{x}_{t-1}\in \mathbb{R}^{C \times H \times W}$ of $\hat{x}_{t-1}$, where $C$ is the channel size, to obtain a temporal predictor $x_{c}\in \mathbb{R}^{C \times H \times W}$. How to use $x_c$ for inter-frame coding affects crucially the trade-off between compression performance and complexity. We shall address this issue by exploring and comparing three coding techniques -- conditional coding, conditional residual coding, and masked conditional residual coding -- based on the same coding components used in DCVC. Section \ref{sec:inter} details how $x_c$ is employed in each of these coding techniques.

\subsection{Inter-frame Coding}
\label{sec:inter}
\subsubsection{Conditional Coding}
DCVC~\cite{dcvc} adopts conditional coding for inter-frame coding. As illustrated in Fig.~\ref{fig:schemes}~(a), the input frame $x_t$ is coded conditionally based on the temporal predictor $x_c$. To mitigate the information bottleneck from using $x_c$ as the condition signal, DCVC employs a two-pronged approach: (1) retaining $x_c$ at full resolution and refining it as $\dot{x}_c$, and (2) choosing a large channel size $C$ (e.g. 64) for $x_c$. As shown in Fig.~\ref{fig:schemes}~(a), $x_t$ and $\dot{x}_c$ are concatenated as the input to the inter-frame codec, $\{G^{enc}, G^{dec}\}$. Likewise, on the decoder side, the decoded signal of dimension ${64 \times H \times W}$ is concatenated with $\dot{x}_c$ and fed into the frame generator to reconstruct the input frame. DCVC~\cite{dcvc} is typical of modern learned video codecs that adopt conditional coding. While it shows very promising coding performance, the need to fetch a large number of full-resolution features in reconstructing a video frame incurs not only high computational complexity but more notably high memory access bandwidth. The latter turns the decoding process into a memory bound operation. This work is aimed at understanding its rate-distortion-complexity trade-off by varying the channel size $C$. 


\subsubsection{Conditional Residual Coding}
As shown in \cite{cond_res_coding}, when the temporal predictor $x_c$ is a good predictor for the coding frame $x_c$ such that $H(x_t-x_c) \leq H(x_t)$ holds true, encoding the prediction residue $x_t-x_c$ with \textcolor{black}{$f(x_c)$} as the condition signal is at least as efficient as encoding the target frame $x_t$ with the same condition \textcolor{black}{$f(x_c)$}. Here $f(x_c)$ represents the processed version of the temporal predictor $x_c$. In this work, $f(x_c)=\dot{x_c}$. However, in \cite{cond_res_coding}, the temporal predictor $x_c$ is a 3-channel signal, whereas $x_c$ in DCVC~\cite{dcvc} comprises C-channel feature maps. As depicted in Fig.~\ref{fig:schemes}~(b), to adapt DCVC~\cite{dcvc} to conditional residual coding, we use a 3x3 convolutional layer to transform $x_c$ into $\ddot{x}_c\in \mathbb{R}^{3 \times H \times W}$, a 3-channel pixel-domain temporal predictor. Since $\dot{x}_c$ is obtained from a refinement network that may introduce information loss, we use the original $x_c$ instead of $\dot{x}_c$ to obtain $\ddot{x}_c$. To encode a coding frame $x_t$, our conditional residual inter-frame codec encodes the residue signal $x_t-\ddot{x_c}$, conditioned on $\dot{x}_c$. 

\subsubsection{Masked Conditional Residual Coding}
While $H(x_t-\ddot{x}_c) \leq H(x_t)$ generally holds true, it can be violated at the sub-picture level. For example, in regions with dis-occlusion or unreliable motion estimates, conditional coding is more efficient than conditional residual coding. To enjoy the merits of both conditional coding and conditional residual coding, Chen~\emph{et~al.}~\cite{maskCRT} propose using a pixel-wise soft mask $m~\in~[0,1]^{1 \times H \times W}$ to blend the target frame $x_t$ and the residual frame $x_t-\ddot{x}_c$ as $(1-m) \odot x_t + m \odot (x_t - \ddot{x}_c)$, where $\odot$ is element-wise multiplication, to serve as the input signal to the conditional inter-frame codec. When $\ddot{x}_t$ forms a good prediction of $x_t$ in some image regions, the corresponding mask values should ideally approach 1. In this case, conditional residual coding is chosen. Conversely, when the prediction $\ddot{x}_t$ is poor, the mask values should approach 0, turning the coding mode into conditional coding. In Fig.~\ref{fig:schemes}~(c), we extend from conditional residual coding in Fig.~\ref{fig:schemes}~(b) to masked conditional residual coding. We follow \cite{maskCRT} to introduce a mask generator to predict a pixel-wise soft mask $m \in \mathbb{R}^{1 \times H \times W}$. The mask generator takes the decoded flow maps $\hat{f}_t$ and the pixel-domain temporal predictor $\ddot{x}_c$ as inputs. The architecture of the mask generator is elaborated in Fig.~\ref{fig:mask_gen}. To transmit a coding frame $x_t$, our masked conditional residual inter-frame codec encodes $x_t-m\odot \ddot{x}_c$ (equivalent to $(1-m) \odot x_t + m \odot (x_t - \ddot{x}_c))$, conditioned on $\dot{x}_c$, where the predicted $m \in \mathbb{R}^{1 \times H \times W}$ is replicated 3 times as $\mathbb{R}^{3 \times H \times W}$ for element-wise multiplication.


\begin{figure*}[t]
    \centering
    \subfigure{
        \centering
        \includegraphics[width=.32\textwidth]{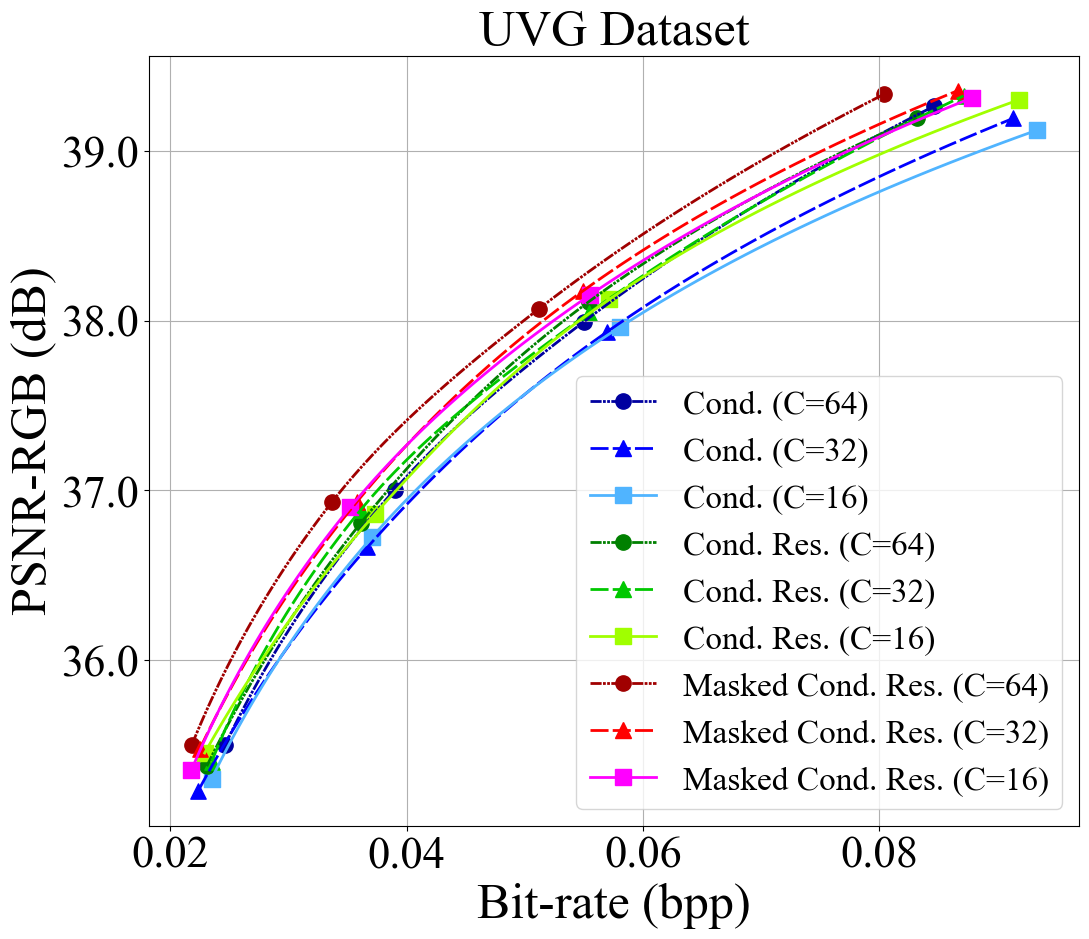}
        \label{fig:rd-a}
        }
    \hspace{-0.35cm}
    \subfigure{
        \centering
        \includegraphics[width=.32\textwidth]{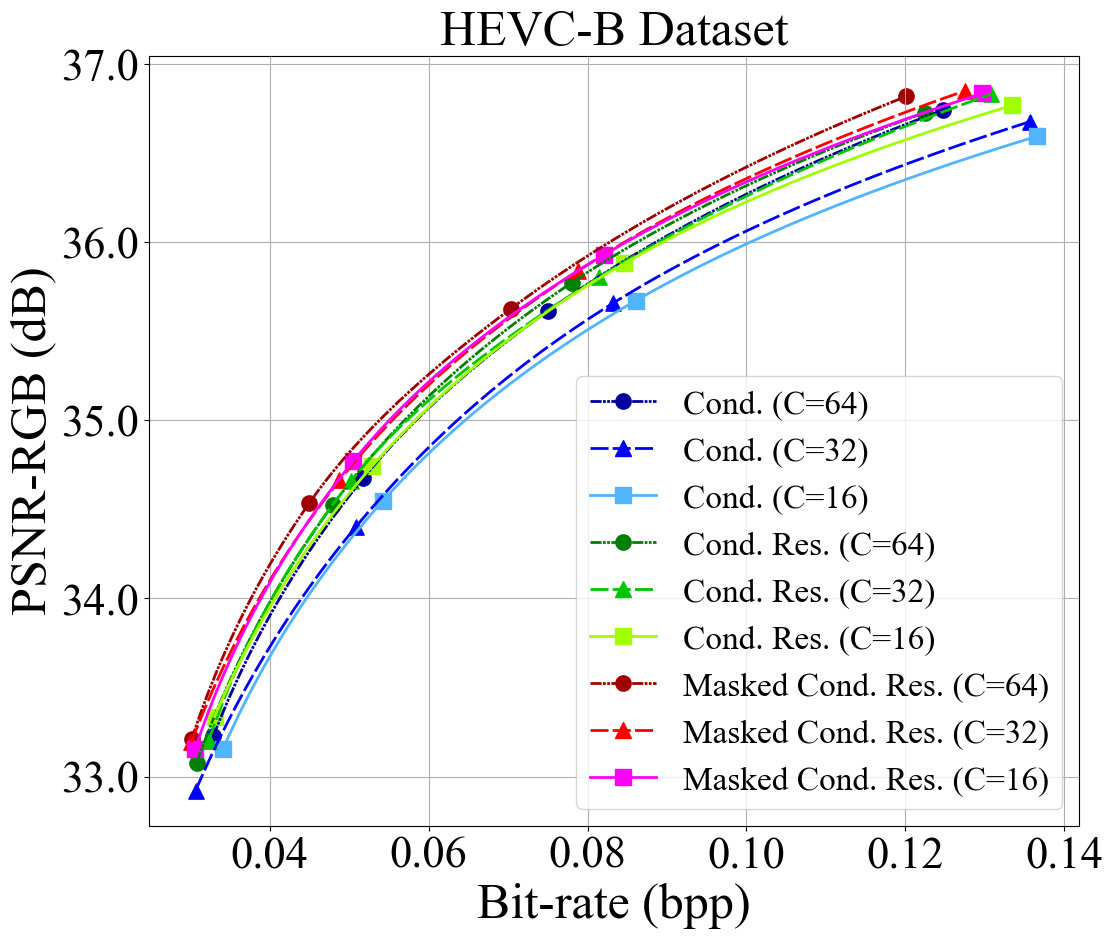}
        \label{fig:rd-b}
    }
    \hspace{-0.35cm}
    \subfigure{
        \centering
        \includegraphics[width=.32\textwidth]{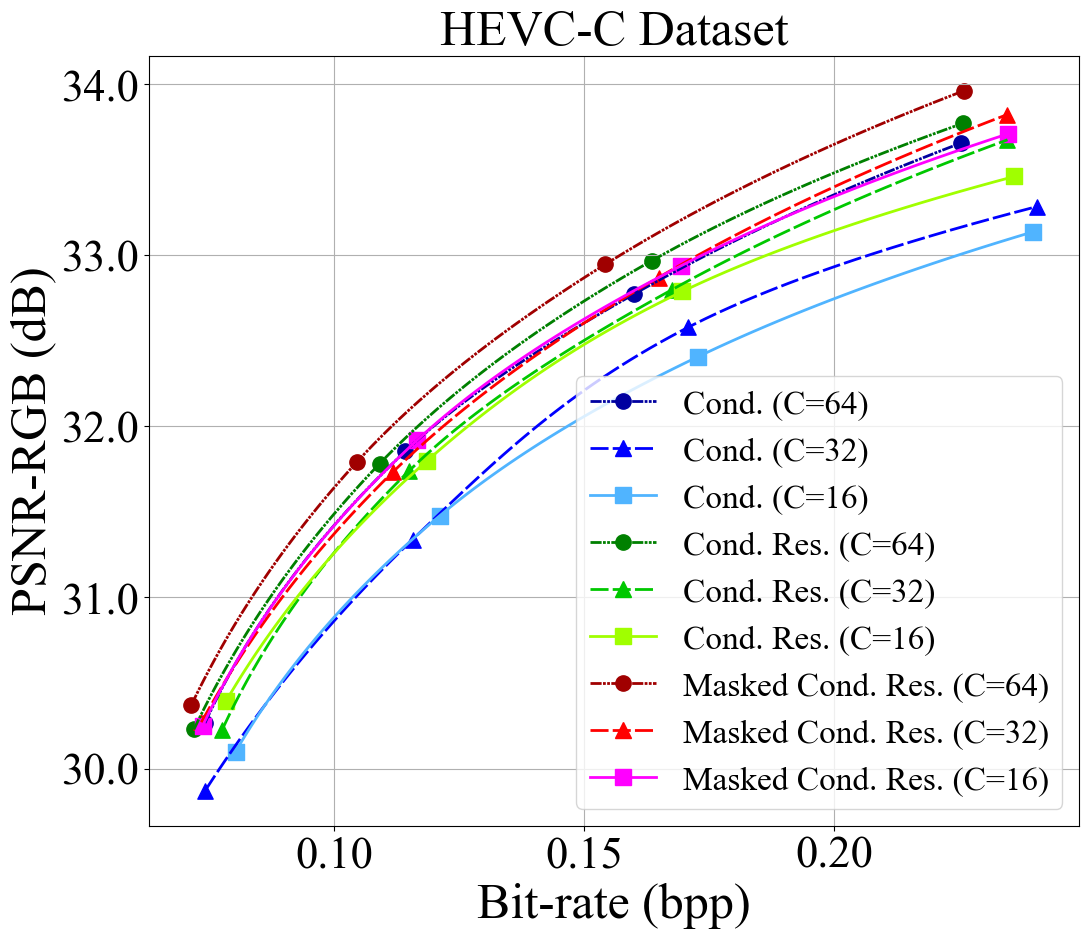}
        \label{fig:rd-c}
    }
    \subfigure{
        \centering
        \includegraphics[width=.32\textwidth]{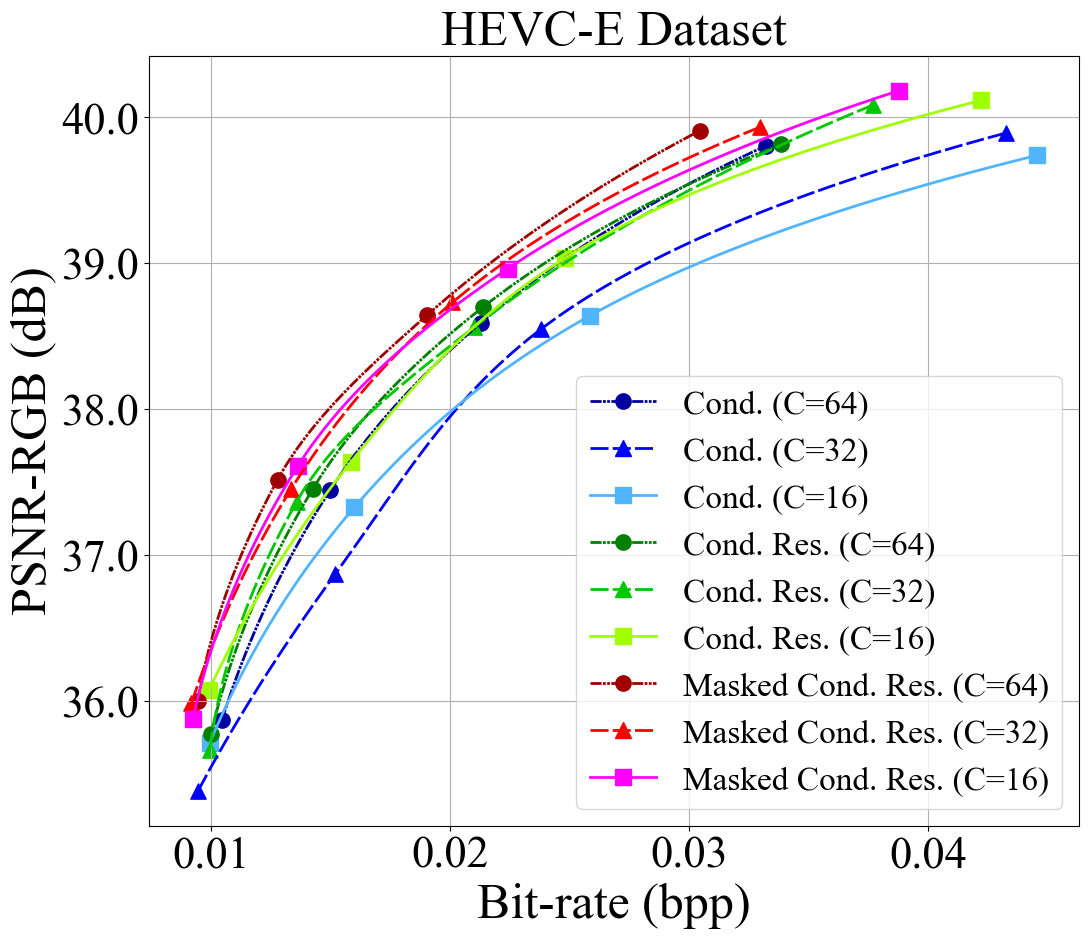}
        \label{fig:rd-d}
        }
    \hspace{-0.35cm}
    \subfigure{
        \centering
        \includegraphics[width=.32\textwidth]{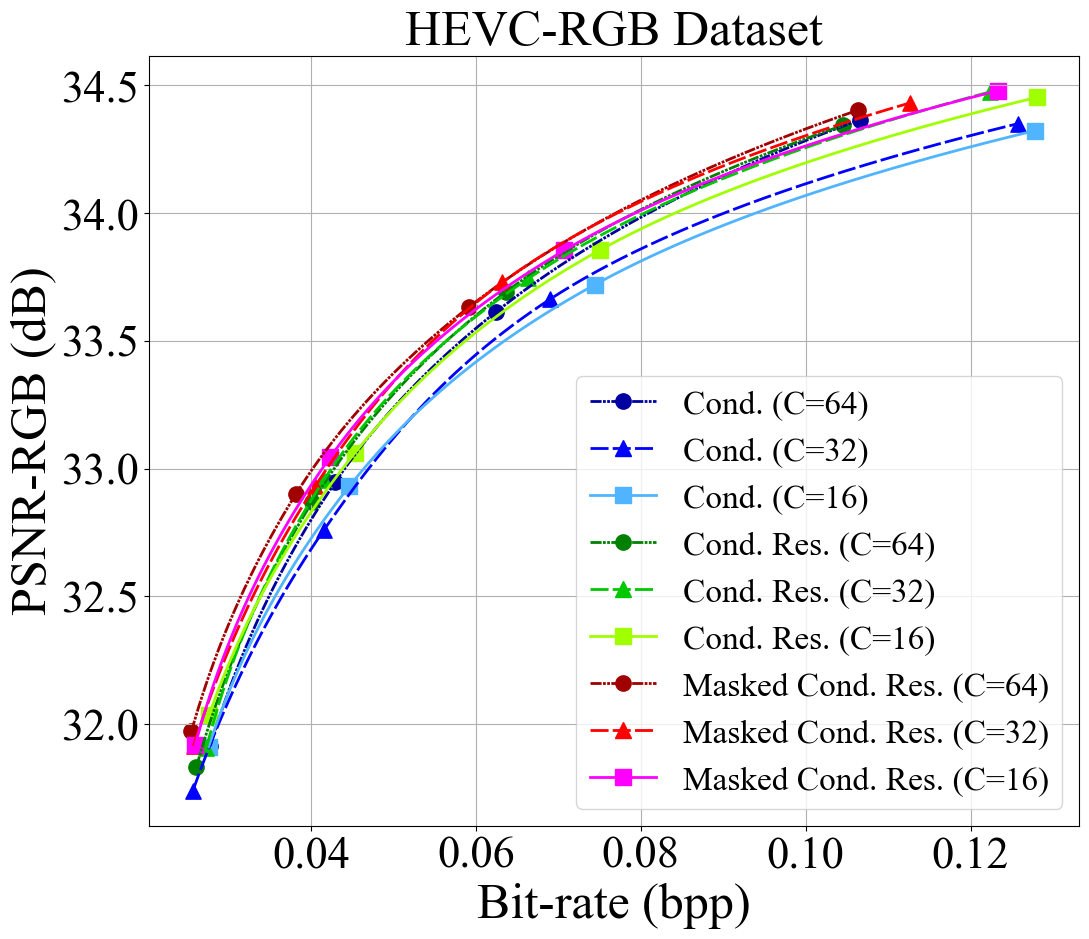}
        \label{fig:rd-e}
    }
    \hspace{-0.35cm}
    \subfigure{
        \centering
        \includegraphics[width=.32\textwidth]{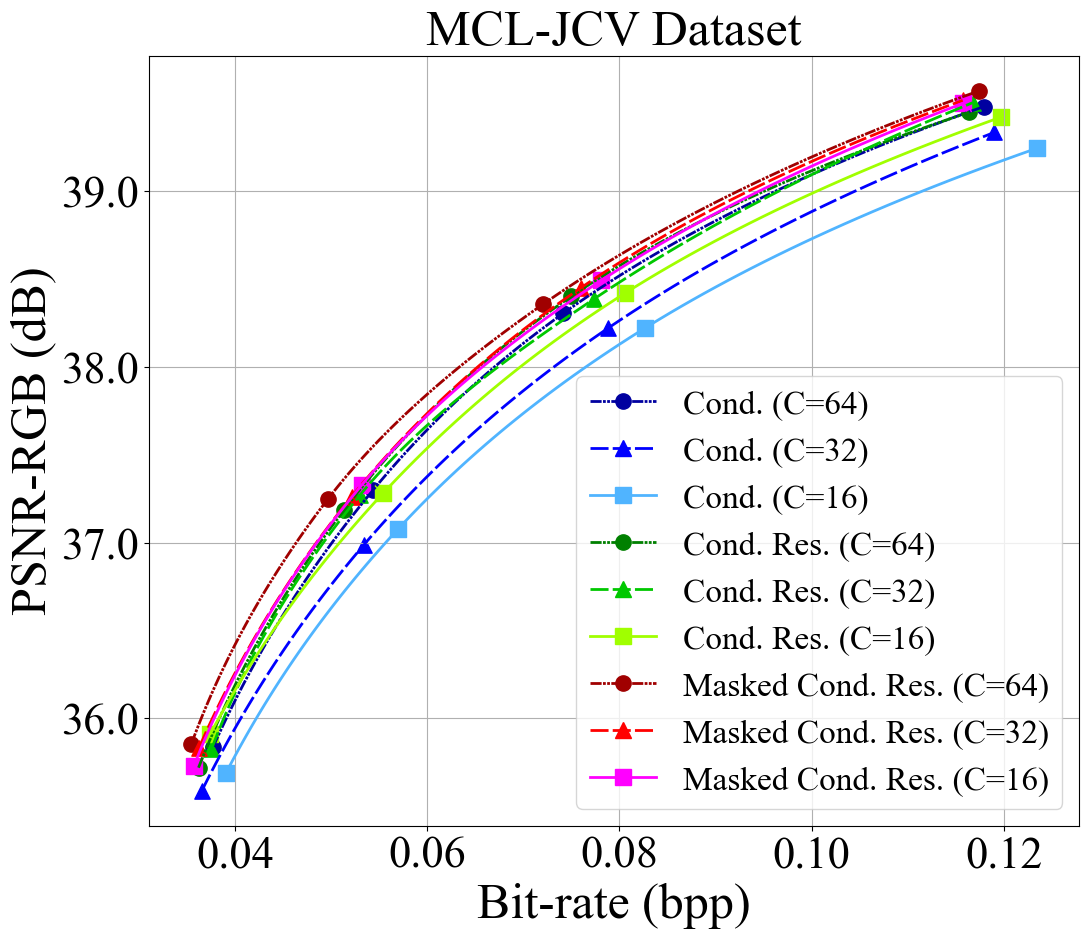}
        \label{fig:rd-f}
    }
    \vspace{-0.1cm}
    \caption{Rate-distortion comparison of conditional coding, conditional residual coding, and masked conditional residual coding under varying bottleneck levels.}
    \vspace{-0.2cm}
    \label{fig:main_RD}
\end{figure*}

\section{Experiments}
\label{sec:experiment}
\subsection{Training} 
We follow the common protocol of learned video compression to train our models on Vimeo-90k dataset~\cite{vimeo}, which contains 91,701 7-frame sequences with resolution $448 \times 256$. We randomly crop the training sequences into $256 \times 256$ patches for training. Our training details are summarized in Table~\ref{table:training}. 
\textcolor{black}{We initialize the conditional coding model with the pre-trained model from~\cite{dcvc_code}, starting training from step 2. Conditional residual coding and masked conditional residual coding models are initialized using the corresponding conditional coding and conditional residual coding models with the same $C$ value, beginning training from step 1.} 
Separate models are trained to optimize the mean-square error (MSE) in the RGB domain with $\lambda=\{256, 512, 1024, 2048\}$.

\begin{table*}[t]
\caption{BD-rate (\%) comparison with the distortion measured in the PSNR-RGB domain. The anchor is conditional coding with $C=64$.}
\label{table:BD_BT709_piece_wise_traditional}
\centering
\begin{tabular}{l|cccccc|c}
\toprule
                         & UVG & HEVC-B & HEVC-C & HEVC-E & HEVC-RGB & MCL-JCV &  Average \\
\midrule                         
Cond. (C=64)             &     0  &     0  &     0  &      0  &     0  &     0  &       0  \\
Cond. Res. (C=64)        &  -1.66 &  -2.57 &  -3.07 &   -3.55 &  -3.14 &  -3.53 &    -2.92 \\
Masked Cond. Res. (C=64) & -11.87 &  -7.66 &  -7.40 &  -13.93 &  -9.68 &  -9.92 &   -10.08 \\
\midrule
Cond. (C=32)             &   6.32 &   7.57 &  16.18 &   15.29 &   4.12 &   5.95 &     9.24 \\
Cond. Res. (C=32)        &  -1.82 &  -1.55 &   3.77 &   -3.49 &  -4.28 &  -2.58 &    -1.66 \\
Masked Cond. Res. (C=32) &  -6.08 &  -5.18 &   0.01 &  -10.73 &  -7.99 &  -5.71 &    -5.95 \\
\midrule
Cond. (C=16)             &   5.72 &   9.73 &  19.79 &   14.08 &   3.76 &  10.31 &    10.57 \\
Cond. Res. (C=16)        &  -0.04 &  -0.61 &   5.39 &   -0.77 &  -3.17 &  -0.30 &     0.08 \\
Masked Cond. Res. (C=16) &  -6.09 &  -5.44 &  -0.28 &  -10.47 &  -8.10 &  -6.91 &    -6.22 \\
\bottomrule
\end{tabular}
\end{table*}


\subsection{Evaluation} 
We evaluate our models on UVG~\cite{uvg}, HEVC Class B~\cite{hevcctc}, HEVC Class C~\cite{hevcctc}, HEVC Class E~\cite{hevcctc}, HEVC-RGB~\cite{hevcrgb} and MCL-JCV~\cite{mcl} datasets. We follow \cite{dcvc_dc} to convert all the test sequences from YUV420 to RGB444 (except for HEVC-RGB, which is already in RGB444 format) using BT.709~\cite{ffmpeg} and provide results for 96-frame encoding with GOP size 32. The reconstructed quality is measured in peak signal-to-noise ratio (PSNR) in the RGB domain and the bit rate in bits-per-pixel (bpp). The average BD-rates over the test sequences in these datasets are reported following the common test protocol of traditional codecs~\cite{hevc,vtm,ecm} by averaging the per-sequence BD-rates. The piecewise cubic interpolation~\cite{interpolation} is used for BD-rate measurements. Positive and negative BD-rate numbers indicate rate inflation and reduction, respectively.


\begin{figure*}[t!]
\centering
\resizebox{\textwidth}{!}
{
\Large
\begin{tabular}{>{\centering\arraybackslash} m{0.2cm} >{\centering\arraybackslash} m{5.4cm} | >{\centering\arraybackslash} m{5.4cm} | >{\centering\arraybackslash} m{5.4cm} >{\centering\arraybackslash} m{5.4cm} >{\centering\arraybackslash} m{5.4cm} }
    &       & Cond. Res.   & \multicolumn{3}{c}{Masked Cond. Res.} \\
    & $x_t$ & $ x_t - \ddot{x_c}$ & $ x_t - m \odot \ddot{x_c}$ & $m \odot \ddot{x_c}$ & $m$ \\
    {\rotatebox{90}{C=64}} &
    \includegraphics[width=0.3\textwidth]{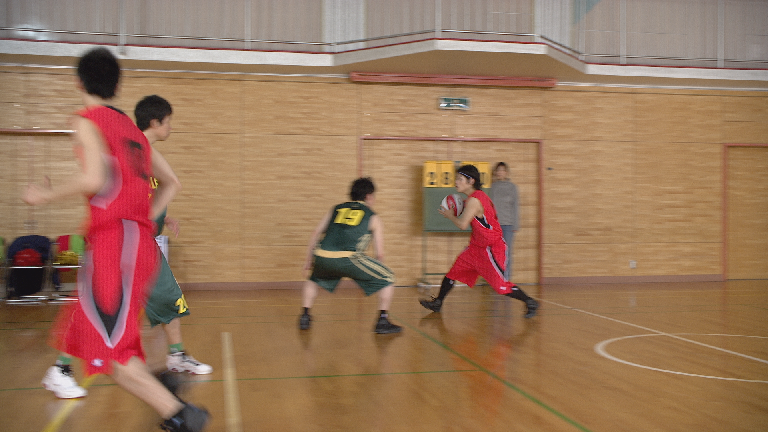}&
    \includegraphics[width=0.3\textwidth]{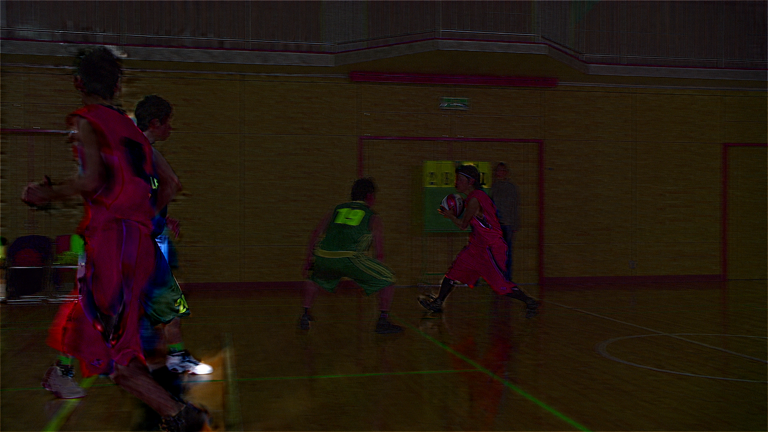}&
    \includegraphics[width=0.3\textwidth]{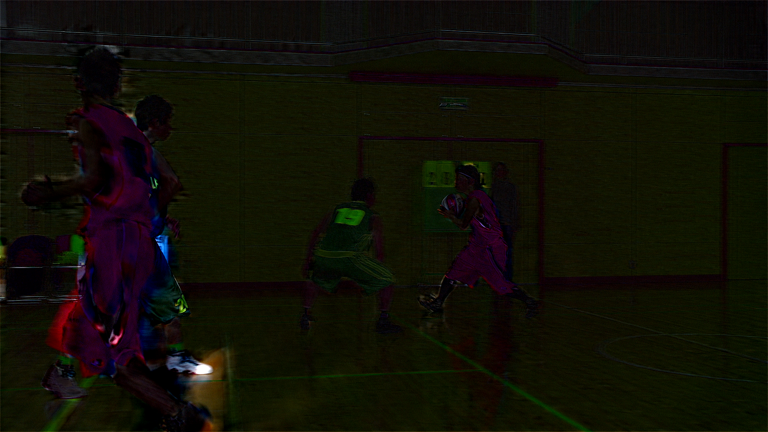}&
    \includegraphics[width=0.3\textwidth]{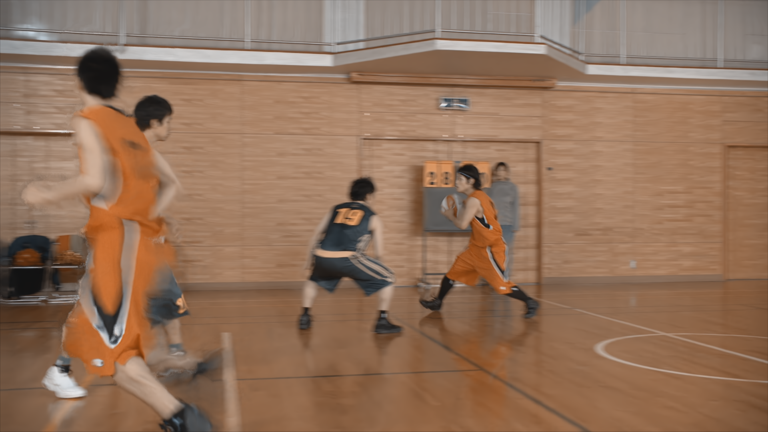}&
    \includegraphics[height=88pt]{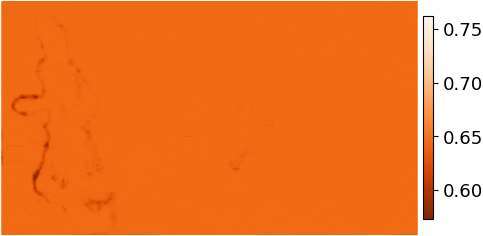} \\
    
    {\rotatebox{90}{C=32}} &
    \includegraphics[width=0.3\textwidth]{Figure/visualization/BasketballDrive_down_04/gt_frame_4_downsampled.png}&
    \includegraphics[width=0.3\textwidth]{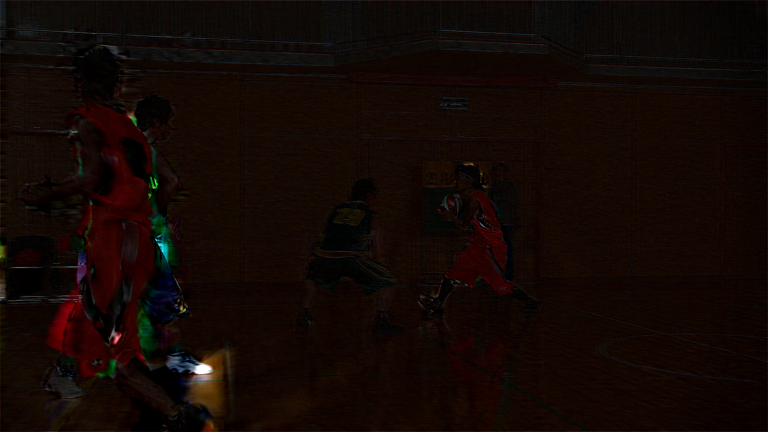}&
    \includegraphics[width=0.3\textwidth]{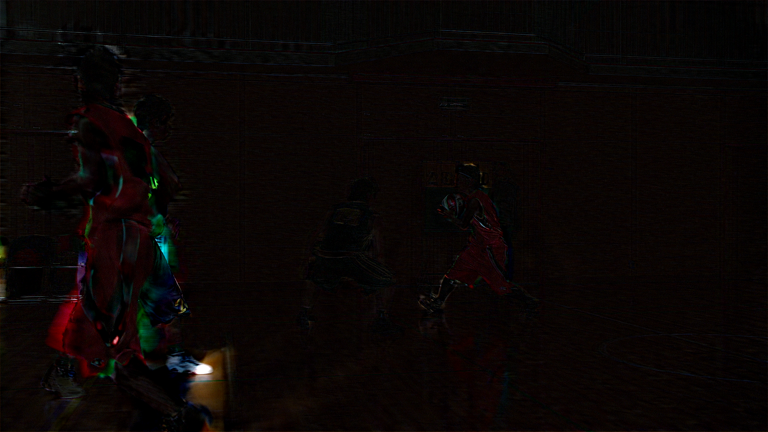}&
    \includegraphics[width=0.3\textwidth]{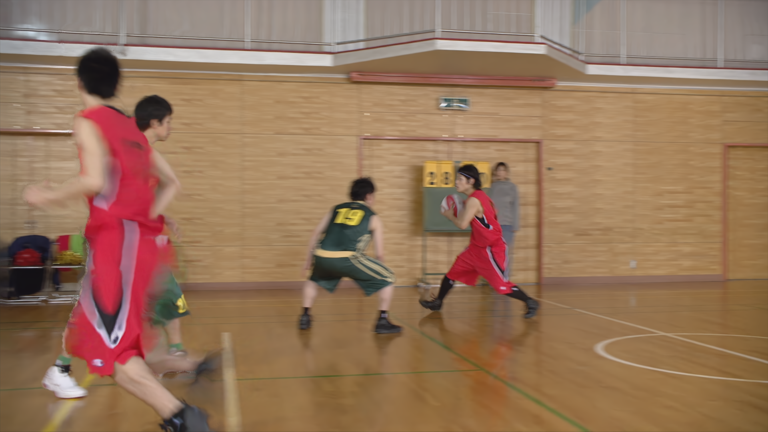}&
    \includegraphics[height=88pt]{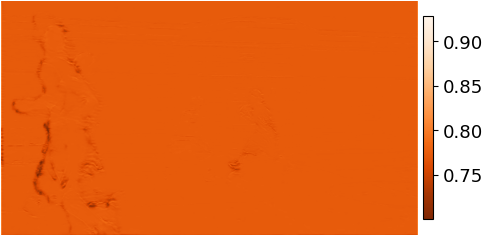} \\

    {\rotatebox{90}{C=16}} &
    \includegraphics[width=0.3\textwidth]{Figure/visualization/BasketballDrive_down_04/gt_frame_4_downsampled.png}&
    \includegraphics[width=0.3\textwidth]{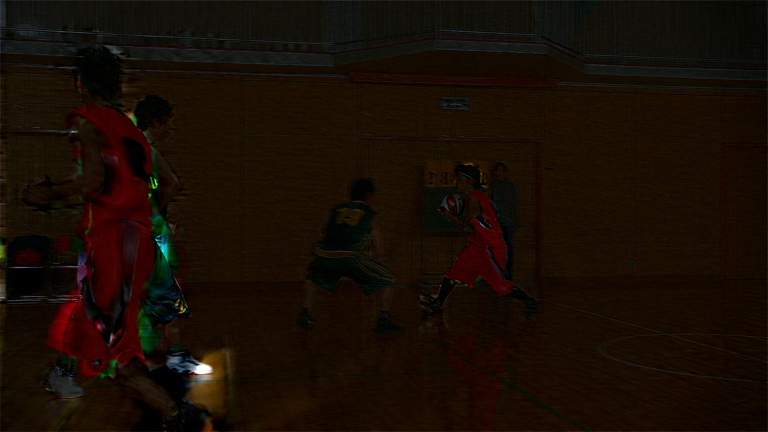}&
    \includegraphics[width=0.3\textwidth]{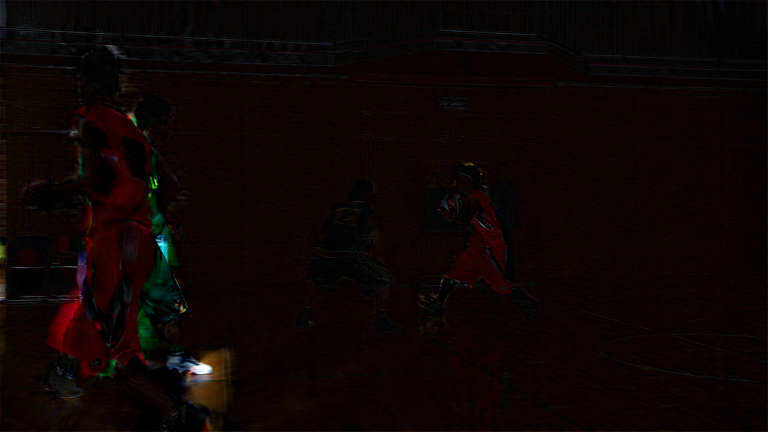}&
    \includegraphics[width=0.3\textwidth]{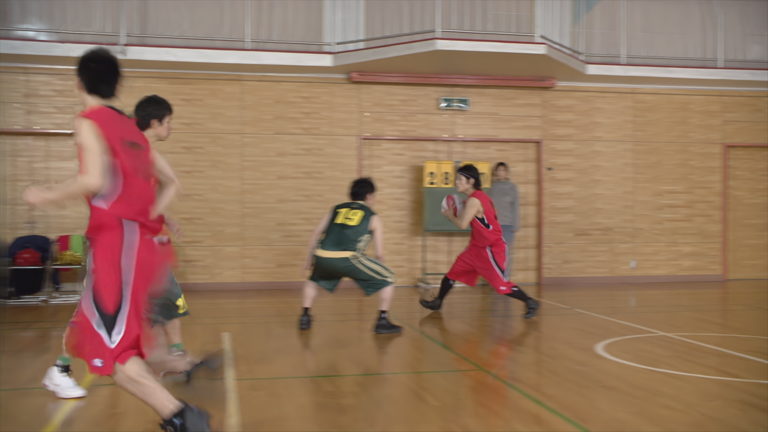}&
    \includegraphics[height=88pt]{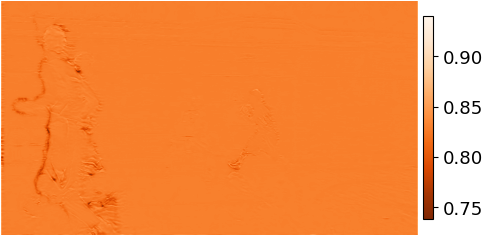} \\

    {\rotatebox{90}{C=64}} &
    \includegraphics[width=0.3\textwidth]{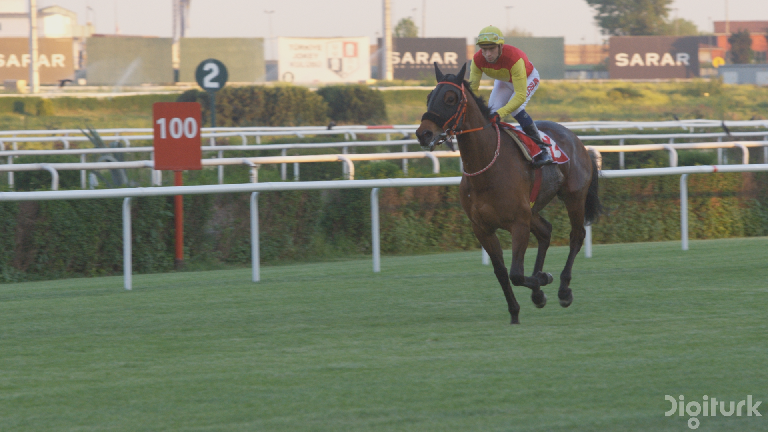}&
    \includegraphics[width=0.3\textwidth]{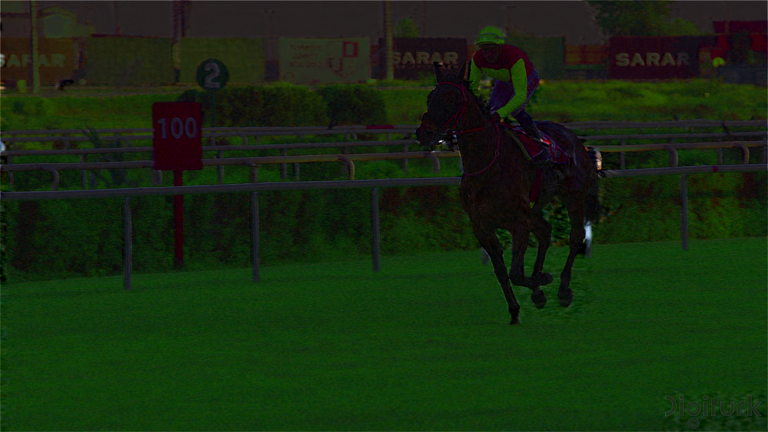}&
    \includegraphics[width=0.3\textwidth]{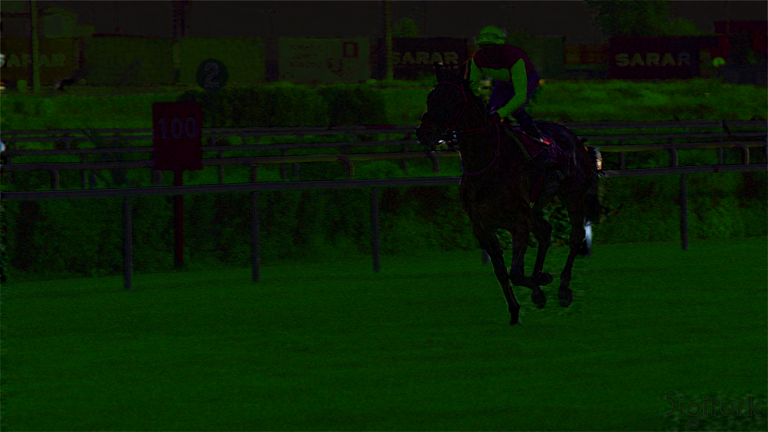}&
    \includegraphics[width=0.3\textwidth]{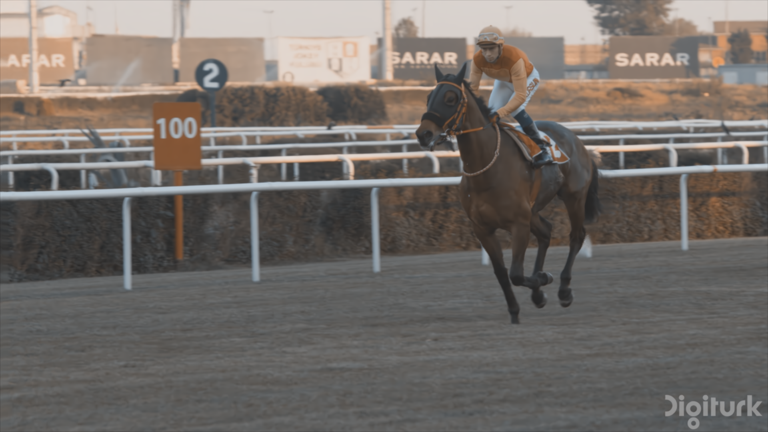}&
    \includegraphics[height=88pt]{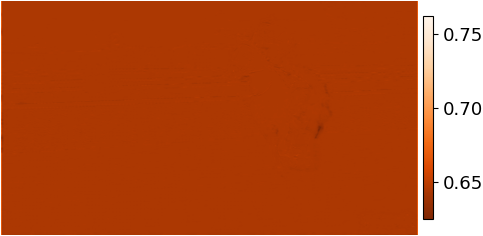} \\
    
    {\rotatebox{90}{C=32}} &
    \includegraphics[width=0.3\textwidth]{Figure/visualization/Jockey_down_04/gt_frame_11_downsampled.png}&
    \includegraphics[width=0.3\textwidth]{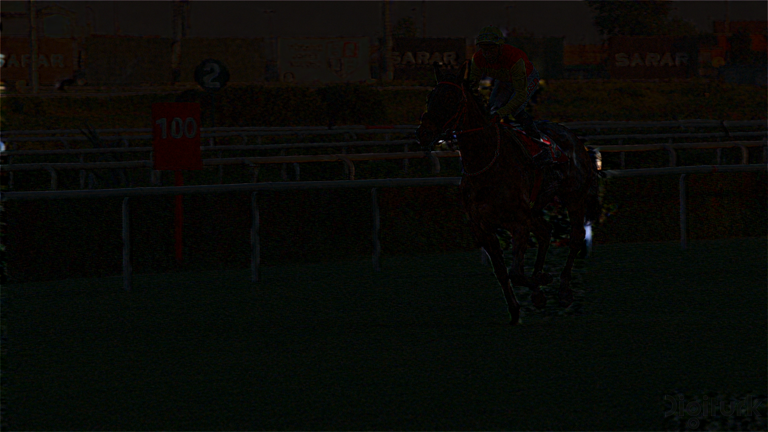}&
    \includegraphics[width=0.3\textwidth]{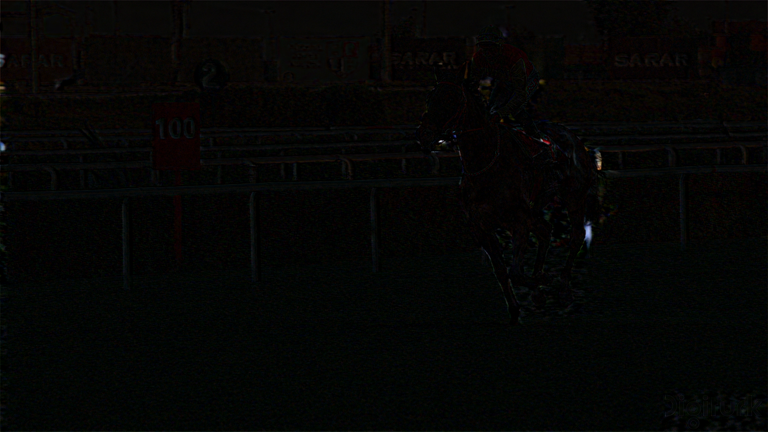}&
    \includegraphics[width=0.3\textwidth]{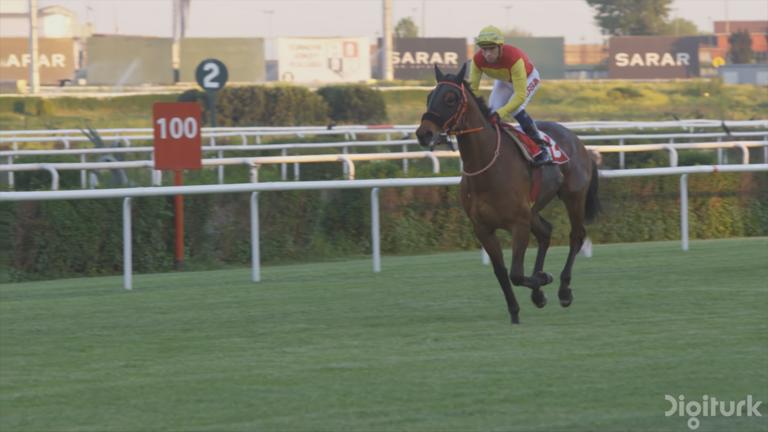}&
    \includegraphics[height=88pt]{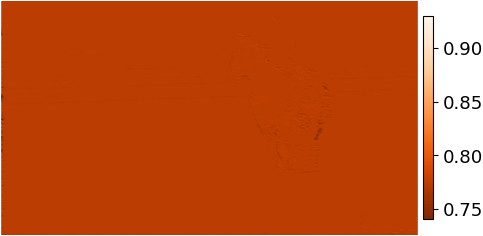} \\

    {\rotatebox{90}{C=16}} &
    \includegraphics[width=0.3\textwidth]{Figure/visualization/Jockey_down_04/gt_frame_11_downsampled.png}&
    \includegraphics[width=0.3\textwidth]{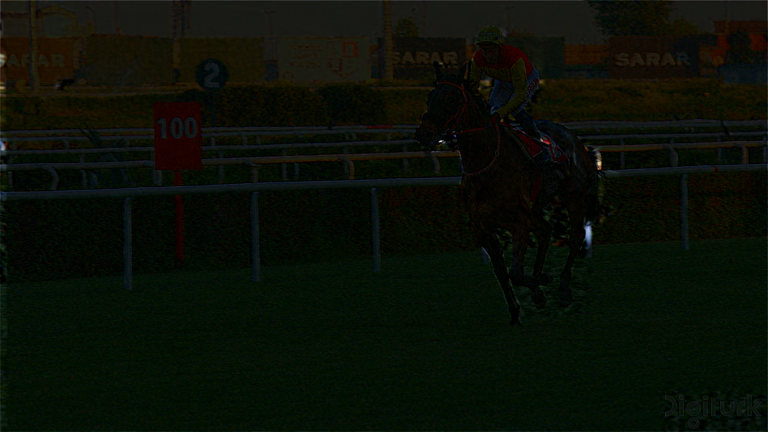}&
    \includegraphics[width=0.3\textwidth]{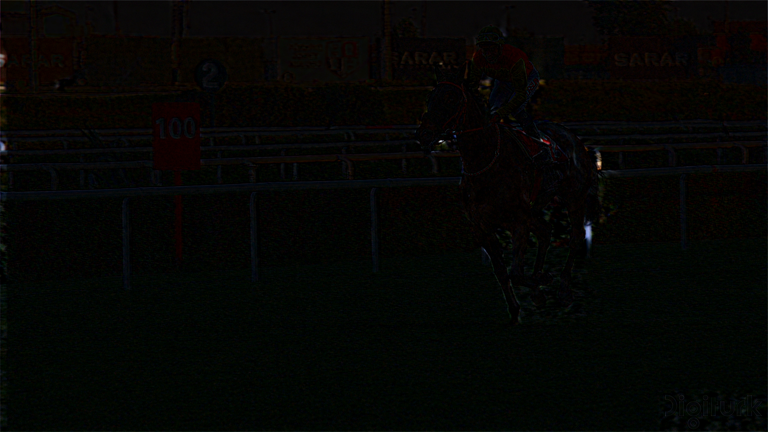}&
    \includegraphics[width=0.3\textwidth]{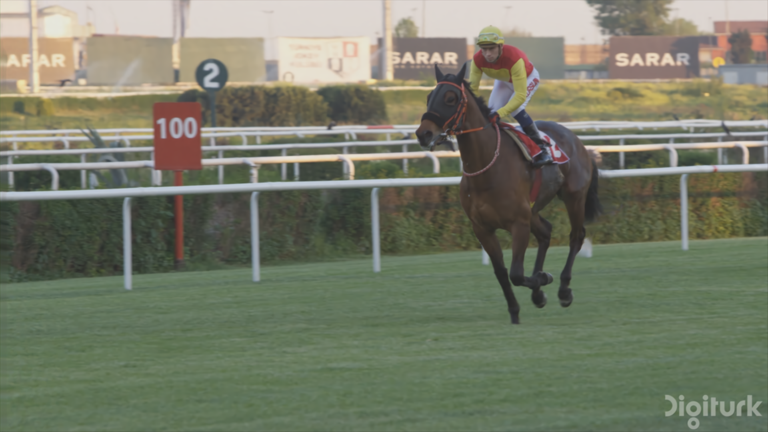}&
    \includegraphics[height=88pt]{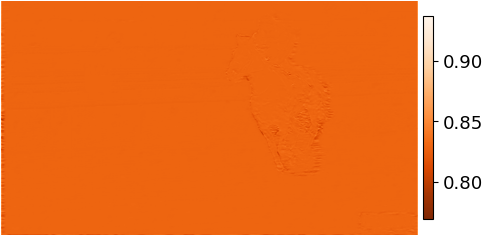} \\

\end{tabular}
}

\caption{Visualization of the input signals and masks for conditional residual coding and masked conditional residual coding under different bottleneck levels. From left to right, the second column shows the input signals for conditional residual coding. The third, fourth, and fifth columns visualize the input signals, masked prediction residues, and masks for masked conditional residual coding, respectively.}

\vspace{-0.05cm}
\label{fig:vis}
\end{figure*}

\begin{table*}[t]
\caption{Comparison of the BD-rate and complexity in terms of the encoding/decoding MACs, model size and the required channel size of the full-resolution condition signal $\dot{x}_c$ for inter-frame coding. The values in parentheses indicate the change in percentage terms relative to conditional coding with $C=64$.}
\vspace{-0.1cm}
\label{table:complexity}
\centering
\begin{tabular}{l|c|cccc}
\toprule
                         & BD-rate (\%) & Encoding (kMACs/pixel) & Decoding (kMACs/pixel) & Model Size (M)  & Channel Size of $\dot{x}_c$ \\ \midrule
Cond. (C=64)             & 0            & 1153                   & 762                    & 7.944           & 64                          \\
Cond. Res. (C=64)        &  -2.92        & 1155 (+0.17\%)          & 764 (+0.26\%)           & 7.946 (+0.03\%) & 64                          \\
Masked Cond. Res. (C=64) & -10.08        & 1212 (+5.12\%)          & 821 (+7.74\%)           & 8.169 (+2.83\%) & 64                          \\ \midrule
Cond. (C=32)             &   9.24        & 970 (-15.87\%)         & 592 (-22.31\%)         & 7.684 (-3.27\%) & 32                          \\
Cond. Res. (C=32)        &  -1.66        & 971 (-15.78\%)         & 593 (-22.17\%)         & 7.685 (-3.26\%) & 32                          \\
Masked Cond. Res. (C=32) &  -5.95        & 1027 (-10.93\%)        & 649 (-14.83\%)         & 7.908 (-0.45\%) & 32                          \\ \midrule
Cond. (C=16)             &  10.57        & 913 (-20.82\%)         & 541 (-29.00\%)         & 7.589 (-4.47\%) & 16                          \\
Cond. Res. (C=16)        &   0.08         & 913 (-20.82\%)         & 541 (-29.00\%)         & 7.589 (-4.47\%) & 16                          \\
Masked Cond. Res. (C=16) &  -6.22        & 970 (-15.87\%)         & 598 (-21.52\%)         & 7.812 (-1.66\%) & 16                          \\
\bottomrule

\end{tabular}
\vspace{-0.3cm}
\end{table*}

\subsection{Rate-Distortion Performance}
Fig.~\ref{fig:main_RD} compares the 
rate-distortion performance of conditional coding, conditional residual coding, and masked conditional residual coding. Table~\ref{table:BD_BT709_piece_wise_traditional} reports their BD-rate savings with respect to conditional coding with $C=64$. We adjust the channel size $C$ of $\tilde{x}_{t-1}, x_c$ and $\dot{x}_c$ to investigate the impact of the information bottleneck on their coding performance. 
From Table~\ref{table:BD_BT709_piece_wise_traditional}, the following observations are immediate: 

(1) Conditional residual coding consistently outperforms conditional coding under the same $C$ values across different datasets. As expected, masked conditional residual coding achieves even higher BD-rate savings. 

(2) Conditional coding is more sensitive to the information bottleneck than both conditional residual coding and masked conditional residual coding. As the information bottleneck becomes more severe (i.e., by reducing $C$ from 64 to 32 or to 16), the coding performance of conditional coding decreases significantly, with the BD-rate inflated by about 9\% for $C=32$ and 10\% for $C=16$. In contrast, the BD-rate inflation of conditional residual coding and masked conditional residual coding is relatively modest when $C$ is reduced.


(3) Conditional residual coding and masked conditional residual coding are able to achieve comparable or even better coding performance than conditional coding with a smaller $C$ value. Specifically, conditional residual coding with $C=16$ achieves comparable coding performance to conditional coding with $C=64$, while conditional residual coding with $C=32$ and masked conditional residual coding with $C=32$ and $C=16$ shows higher coding gain than conditional coding with $C=64$. A smaller $C$ value generally implies lower complexity. An in-depth rate-distortion-complexity analysis is provided in Section~\ref{sec:complexity}.

To further investigate the inner workings of conditional residual coding and masked conditional residual coding under different bottleneck levels, Fig.~\ref{fig:vis} visualizes their input signals and the masks. We see that when the bottleneck issue is mild ($C=64$), the resulting input signals of both codecs has more content information. In a sense, both codecs appear to operate in the conditional coding mode. In contrast, when $C=32$ or $16$, the input signals resemble more closely a form of residual signals. This observation is in line with the findings in \cite{cond_res_coding}, which indicate that when the bottleneck issue is mild, conditional residual coding and conditional coding perform comparably to each other. Additionally, it is worth noting that with masked conditional residual coding, the masks $m$ exhibit relatively lower values in regions where motion estimates are unreliable, such as regions with dis-occlusion, around object boundaries or with complex motion. In other words, conditional coding is more preferred in these regions.


\subsection{Complexity Analysis}
\label{sec:complexity}
Table~\ref{table:complexity} and Fig.~\ref{fig:teaser} report the rate-distortion-complexity trade-offs of conditional coding, conditional residual coding, and masked conditional residual coding under different levels of the information bottleneck. As shown, conditional residual coding and masked conditional residual coding achieve higher coding performance while requiring lower complexity. For example, when comparing conditional coding with $C=64$, conditional residual coding with $C=32$, and masked conditional residual coding with $C=16$, conditional residual coding achieves a 1.66\% BD-rate saving to conditional coding while masked conditional residual coding has a 6.22\% BD-rate saving. Notably, in this comparison, both conditional residual coding and masked conditional residual coding exhibit approximately 16\% and 22\% reductions in encoding and decoding kMACs/pixel, respectively, as compared to conditional coding. In contrast, at a similar complexity level (about 16\% and 22\% reductions in encoding and decoding kMACs/pixel), the coding performance of conditional coding with $C=32$ drops by about 9\%. The channel size $C$ also has an impact on the memory footprint of these coding schemes. For similar coding performance, both conditional residual coding and masked conditional residual coding require fewer channels than conditional coding. 

\section{Conclusion}
\label{sec:conclusion}
This work explores the rate-distortion-complexity trade-offs of neural video codecs that adopt conditional coding, conditional residual coding and masked conditional residual coding. Our major findings include (1) conditional residual coding and masked conditional residual coding are able to effectively mitigate the information bottleneck issue of conditional coding, (2) at a similar or even better coding performance level, conditional residual coding and masked conditional residual coding require 84\% and 78\% of the encoding and decoding kMACs/pixel compared to conditional coding. This work paves the way for more efficient video compression techniques with improved performance and reduced computational complexity.

\bibliography{egbib}
\bibliographystyle{IEEEtran}

\end{document}